\documentstyle[aps,epsf,rotating,amssymb]{revtex}

\begin{document}

\title{Off-diagonal disorder in the Anderson model of localization}

\author{P.\ Biswas$^1$, P.\ Cain$^2$, R.\ A.\ R\"omer$^2$, and M.\ 
  Schreiber$^2$}

\address{$^1$ S N Bose National Center For Basic Sciences, Salt Lake
  City 700 091, Calcutta, India\\ Email: ppb@boson.bose.res.in; Tel.:
  +91333353057; Fax: +91333353477\\ $^2$ Institut f\"ur Physik,
  Technische Universit\"at, D-09107 Chemnitz, Germany\\ 
  Email: cain@physik.tu-chemnitz.de; Tel.: +493715313561; Fax:
  +493715313151} 

\date{\today} \maketitle

\begin{abstract}
  We examine the localization properties of the Anderson Hamiltonian
  with additional off-diagonal disorder using the transfer-matrix
  method and finite-size scaling. We compute the localization lengths
  and study the metal-insulator transition (MIT) as a function of
  diagonal disorder, as well as its energy dependence. Furthermore we
  investigate the different influence of odd and even system sizes on
  the localization properties in quasi one-dimensional systems.
  Applying the finite-size scaling approach in conjunction with a
  nonlinear fitting procedure yields the critical parameters of the
  MIT.  In three dimensions, we find that the resulting critical
  exponent of the localization length agrees with the exponent for the
  Anderson model with pure diagonal disorder.
\end{abstract}

\noindent{\bf Introduction.}
The electron localization properties of disordered solids are of high
interest both experimentally and theoretically. A simple approach to
this problem is given by the Anderson model of localization. In this
model one considers a single electron on a lattice with $N$ sites
described by the Hamiltonian
\begin{equation}
\label{eq-hamil}
  H = \sum_{i \neq j}^{N} t_{ij} |i \rangle \langle j| +
      \sum_i^N \epsilon_i |i \rangle \langle i| ,
\end{equation}
where $|i\rangle$ denotes the Wannier states at site $i$.  Disorder is
usually introduced by varying the onsite potential energies
$\epsilon_i$ randomly in the interval $[-W/2,W/2]$.  In this work we
report on the effects of additional off-diagonal disorder
\cite{SouWGE82,Zim82,InuTA94}, i.e., with random hopping elements
$t_{ij} \in [ c - w/2, c + w/2 ]$ between nearest neighbor sites.
Thus $c$ represents the center and $w$ the width of the uniform
off-diagonal disorder distribution. The energy scale is set by keeping
$w=1$ fixed.  This kind of disorder is similar to the random flux
model \cite{Bat97,BatSZK96}, but here the modulus of the $t_{ij}$ is
random while the phase is constant. The Hamiltonian (\ref{eq-hamil})
has a chiral symmetry on a bipartite lattice, i.e., under the
operation $\psi_n \rightarrow (-1)^{n} \psi_n $, the Hamiltonian $H$
changes it sign and as a consequence, if $\epsilon_n$ is an eigenvalue
of $H$, then so is $-\epsilon_n$.

We calculate the localization length $\lambda$ of the electronic wave
function by the transfer-matrix method (TMM), which is based on an
iterative reformulation \cite{KraM93} of Eq.\ (\ref{eq-hamil}) as
\begin{eqnarray}
  \left( \begin{array}{l} \psi_{n+1} \\ \psi_{n} \end{array} \right) 
=
T_n
\left( \begin{array}{l} \psi_{n} \\ \psi_{n-1} \end{array} \right)
=
\left( \begin{array}{cc}
{\bf t}_{n+1}^{-1}(E {\bf 1} - {\bf H}_\perp) & -{\bf t}_{n+1}^{-1}{\bf t}_{n} \\
{\bf 1}                                & {\bf 0}
\end{array}
\right)
\left( \begin{array}{l} \psi_{n} \\ \psi_{n-1} \end{array} \right)
\quad ,
\end{eqnarray}
where $E$ is the energy, ${\bf H}_\perp$ is the Hamiltonian in the
$n$th slice of the TMM bar, ${\bf 1}$ and ${\bf 0}$ are unit and zero
matrix, respectively \cite{MacK83}, and the diagonal matrix ${\bf
  t}_{n}$ represents the hopping elements connecting the $n$th with
the $(n-1)$th slice.  The cross section of the bar is $1$, $M$, and
$M^2$ for spatial dimension $d=1, 2$, and $3$, respectively, and we
always assume periodic boundary conditions for the TMM. Starting with
two initial states $(\psi_1,\psi_0)$ one iterates along the quasi 1D
bar until the desired accuracy is obtained.  The localization length
$\lambda_{d} = 1/\gamma_{\rm min}$ is computed from the smallest
Lyapunov exponent $\gamma_{\rm min}$ of the eigenvalues $\exp (\pm
\gamma_{i})$ of $\lim_{n\rightarrow\infty}(\tau_{n}^\dagger
\tau_{n})^{1/2n}$ with $\tau_{n} = T_{n} T_{n-1} \cdots T_{2} T_{1}$.
Assuming the validity of one-pa\-rameter scaling close to the MIT, we
expect the reduced localization lengths $\lambda_{d}(M)/M$ to
scale onto a scaling curve $\lambda_{d}(M)/M = f_{d}(\xi/M)$
with scaling parameter $\xi$.

\begin{figure}
\epsfxsize=\textwidth
\epsfbox{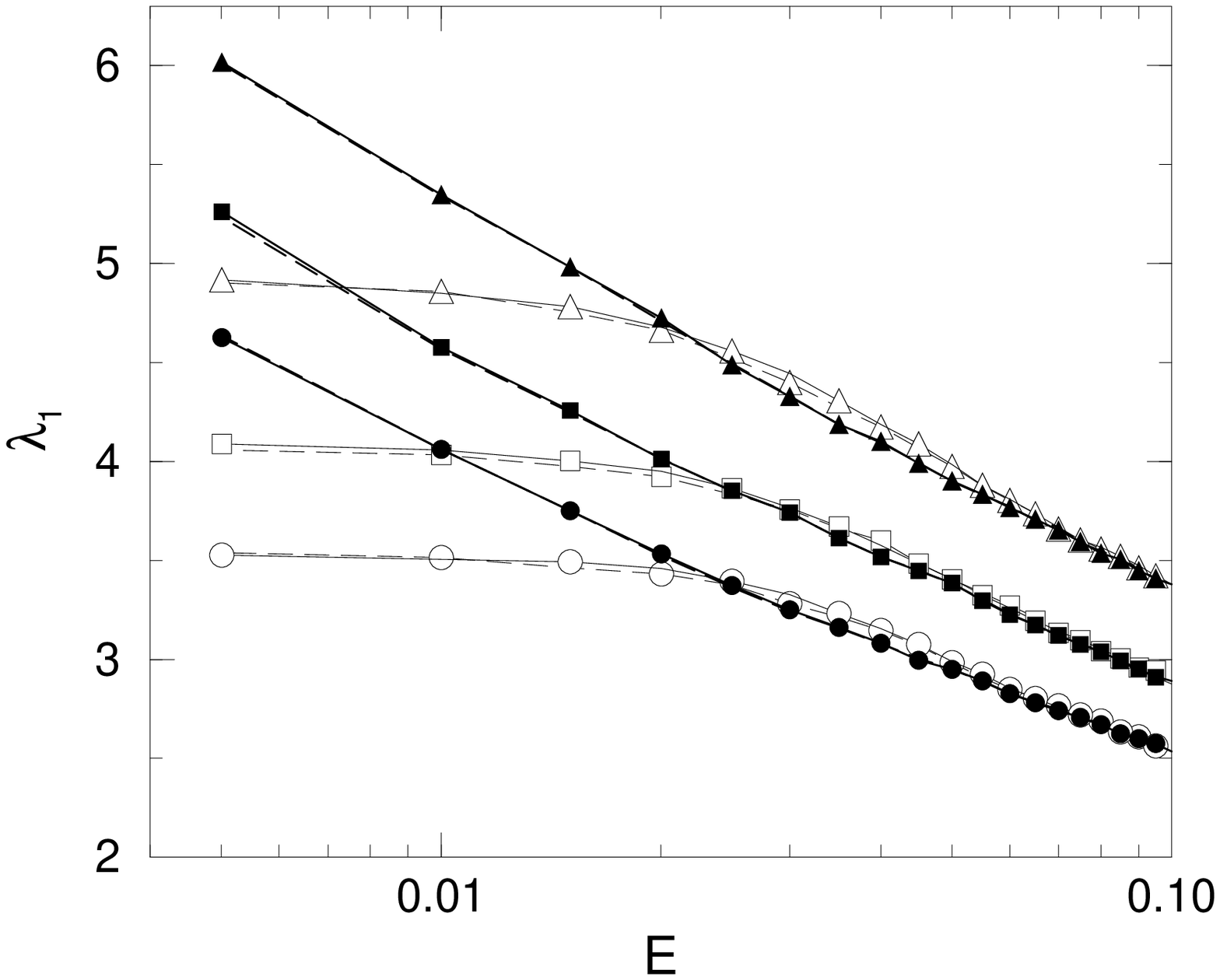}
  \caption{\label{fig-1D-lnEdep}\small 
    Localization length $\lambda_{\rm 1}$ as function of energy $E$
    with off-diagonal disorders $c= 0$ ($\Box$, $\blacksquare$),
    $0.25$ ($\circ$, $\bullet$), and $0.5$ ($\triangle$,
    $\blacktriangle$) for diagonal disorder $W=0.01$ (filled symbols)
    and $0.1$ (open symbols) obtained by TMM with $1\%$ accuracy.
    Solid (dashed) lines indicate data for $E>0$ ($E<0$).}
\end{figure}

\noindent{\bf Localization lengths at the 1D Dyson singularity.}
In 1D it has been shown that the density of states (DOS) of the
disordered model diverges at $E=0$ \cite{Dys53,EggR78}, if $W=0$ on a
bipartite lattice \cite{InuTA94}. This singularity is
intimately related to a divergence of the localization length
\cite{InuTA94,BroMF99}.  Assuming that the localization
properties of the wave function can be described as usual by an
exponential decay, one finds analytically \cite{Mck96}
\begin{equation}
 \lambda_{\rm 1}(E) = \frac{v_{F}}{D} \log\left|\frac{D}{E}\right|
\label{eq-1Dloc}
\end{equation}
with $v_{F}$ the Fermi velocity and $D$ an energy parameter depending
on the strength of the disorder.  In Fig.\ \ref{fig-1D-lnEdep} we show
that this energy dependence is convincingly reproduced in numerical
data of the model: $\lambda_{\rm 1}$ diverges at $E=0$ as in Eq.\ 
(\protect\ref{eq-1Dloc}) for $W\rightarrow 0$. Note that the smallest
values for the localization length are obtained for $c=0.25$. We
remark that the validity of the exponential form of the wave function
has been questioned in previous studies \cite{InuTA94,KozMDR98}
presenting evidence for power-law localization usually related to
critical states.

\noindent{\bf Odd/even effects in quasi 1D systems.}
In 2D, the DOS also has a sharp peak at the band center
\cite{SouWGE82,Zim82,InuTA94} and it was shown by the TMM that the
localization length again diverges \cite{EilRS98a,EilRS98b}. A
finite-size-scaling (FSS) analysis of the TMM data together with
studies of the participation numbers and multifractal properties of
the eigenstates revealed \cite{EilRS98a,EilRS98b} that the states at
$E=0$ show critical behavior like in the 3D Anderson model at the MIT
\cite{GruS95}.

\begin{figure}
\epsfxsize=\textwidth
\epsfbox{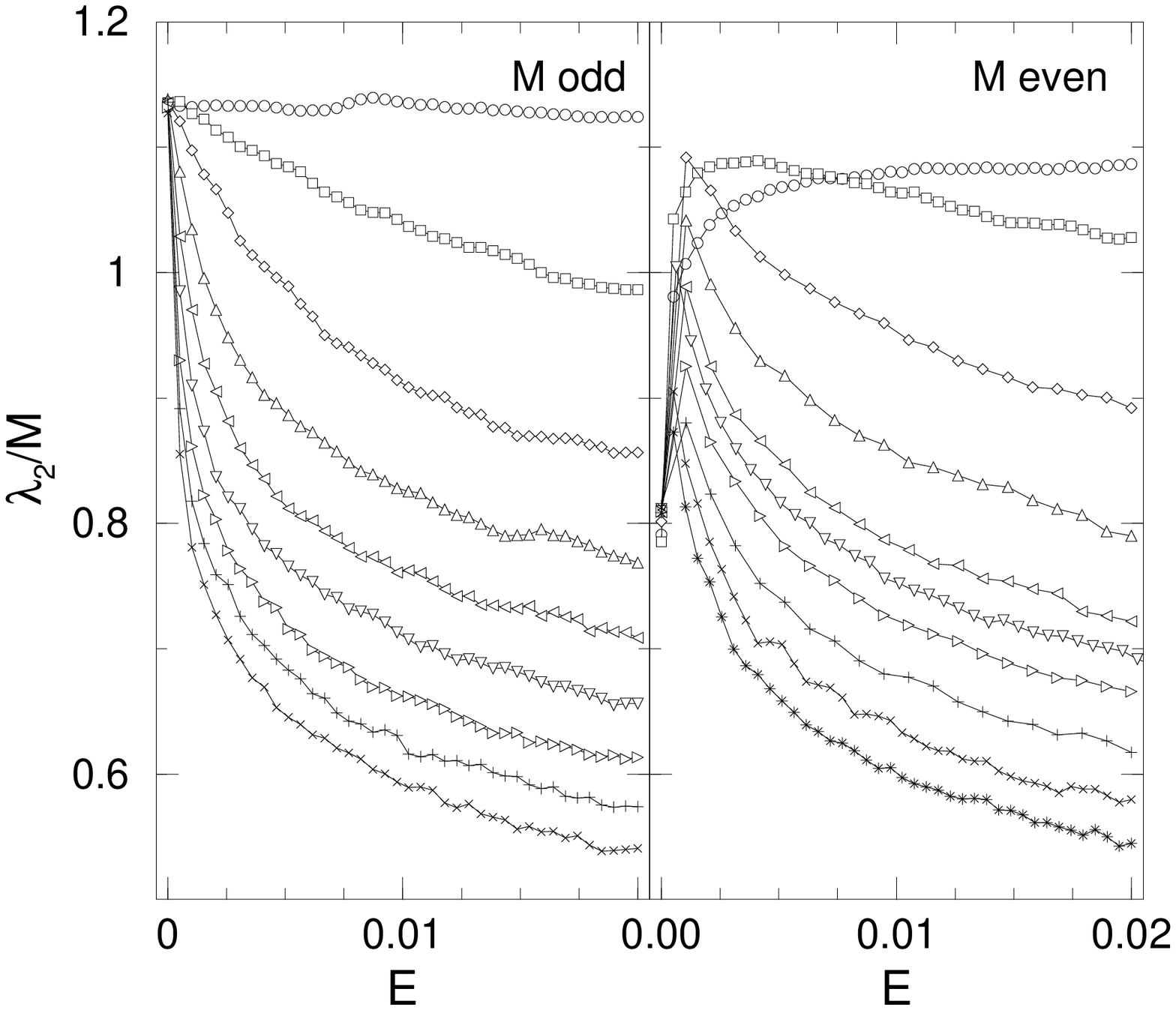}
\caption{
\label{fig-ll}
Reduced localization lengths $\lambda_{\rm 2}/M$ as a function of
energy for odd system sizes (left panel, $M= 3, 7, \ldots, 35$ from
top to bottom) and even system sizes (right panel, $M= 4, 6, 10, 14,
18, 20, 22, 26, 30, 34$ from top to bottom) with random hopping for
$c=0.25$. Note the strong decrease in $\lambda_{\rm 2}/M$ at $E=0$ for
the bipartite even case.}
\end{figure}

\begin{figure}
\epsfxsize=\textwidth
{\epsfbox{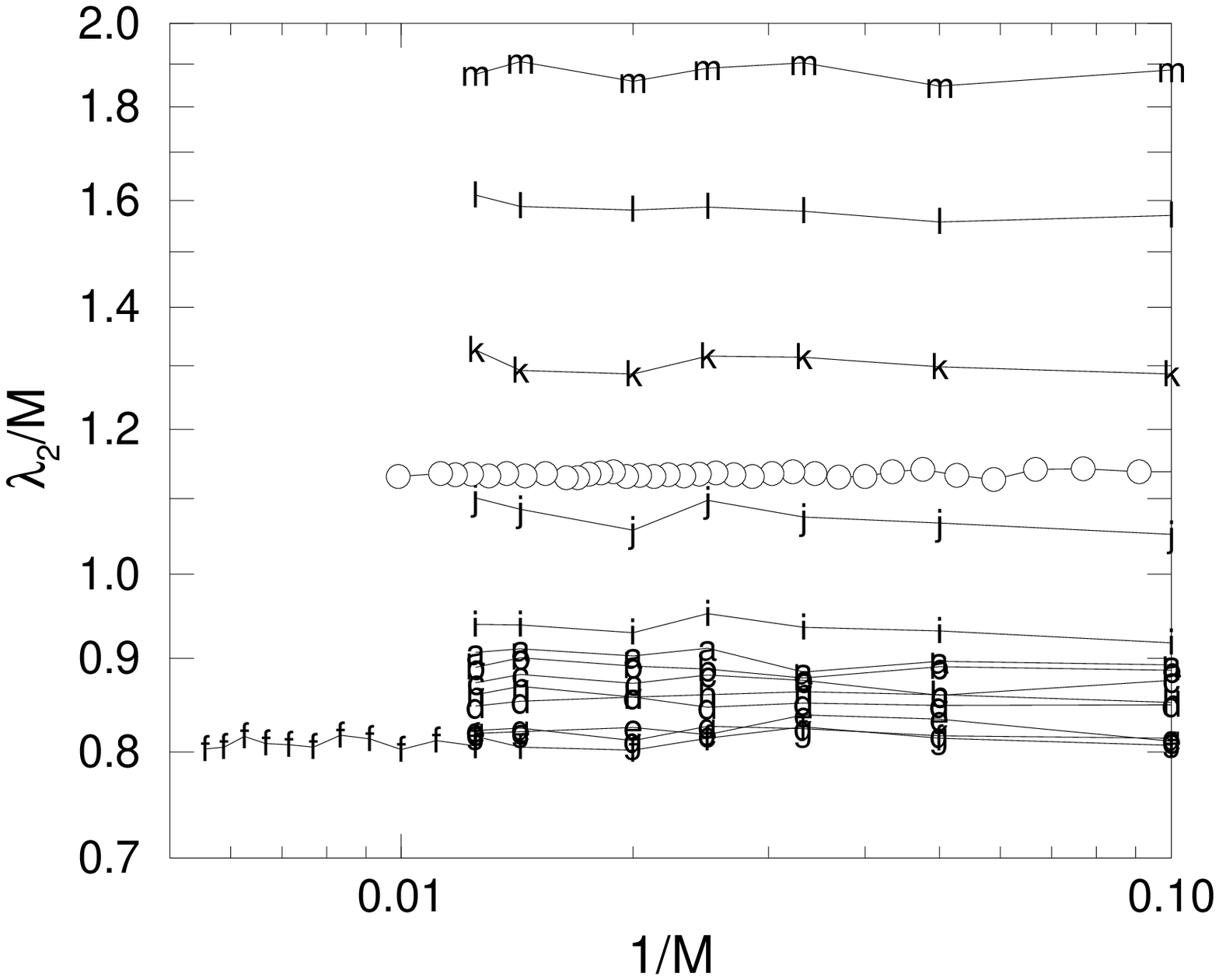}}
\caption{
\label{fig-lam}
Reduced localization length $\lambda_{\rm 2}/M$ vs.\ $1/M$ for $E=0$
and {\em even} $M$ up to $180$ with $1\%$ accuracy corresponding to
$c= 0, 0.05, \ldots, 0.25, \ldots 0.6$ as indicated by a, b, \ldots,
f, \ldots, m \protect\cite{EilRS98a}. The circles represent data for
{\em odd} system sizes $M=11, 13, \ldots, 89$ and $M=101$ with $c=
0.25$, i.e., the smallest $\lambda_{\rm 2}/M$ values, and $0.5\%$
accuracy.}
\end{figure}

Renewed interest in the study of quasi 1D systems with off-diagonal
disorder stems from the fact that the functional form of the
divergence depends on whether $M$ is odd or even as shown by a
Fokker-Planck approach in Ref.\ \cite{BroMF99}. This
question is intimately linked to the presence or absence of
bipartiteness for the given lattice and boundary conditions. In Ref.\ 
\cite{BroMF99} numerical data in support of the analytical
results had already been shown. Here we will broaden the investigation
by studying how the localization length at $E=0$ is influenced by the
odd/even effects.  Eilmes et al.\ \cite{EilRS98a,EilRS98b} observed
that the reduced localization length $\lambda_{\rm 2}/M$ for the 2D
Anderson model with random hopping is independent of the system sizes
at $E=0$ for bipartite lattices up to $M=200$.  Motivated by the
results of Refs.\ \cite{InuTA94,BroMF99}, we show the
behavior of the localization lengths for non-bipartite systems with
odd $M$ as well as bipartite systems with even $M$ in Figs.\ 
\ref{fig-ll} and \ref{fig-lam}.  We see that the values of the reduced
localization length differ for odd and even $M$. Nevertheless,
$\lambda_{\rm 2}/M$ at $E=0$ remains constant and thus critical in the
sense of Ref.\ \cite{EilRS98a} for odd and even $M$ up to $M=101$ and
$180$, respectively.

\begin{figure}
\epsfxsize=0.59\textwidth 
\epsfbox{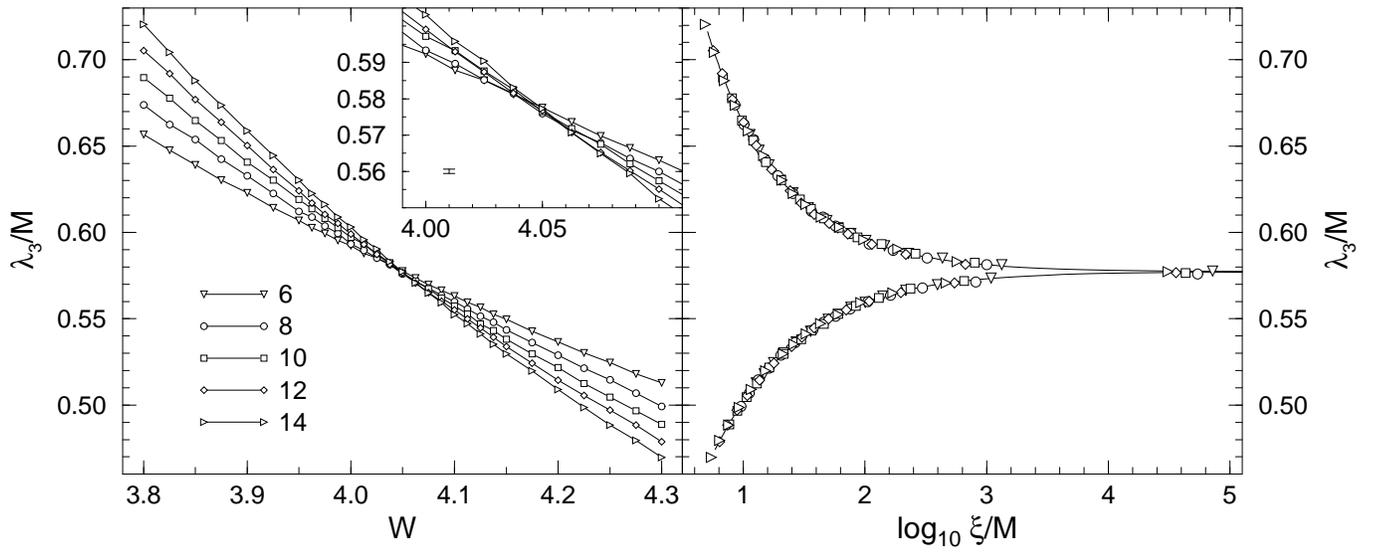}
  \caption{\label{fig-fss}\small 
    Left: Reduced localization length $\lambda_{\rm 3}/M$ as function
    of diagonal disorder $W$ at $c=0$ and $E=0$ obtained by TMM with
    $0.1\%$ accuracy.  Note the MIT at $W_c\approx 4.05$. Inset:
    Enlarged region close to the MIT, the error bar is shown
    separately. Right: FSS curve for the TMM data from the left panel.
    }
\end{figure}

\noindent{\bf Critical exponents in 3D.}
For the 3D model we computed the critical parameters at the MIT when
either $E$ or $W$ are varied across the transition, i.e., $\xi \propto
|E-E_c|^{-\nu_E}$ or $\xi\propto |W-W_c|^{-\nu_W}$ \cite{CaiRS99} at
fixed $c$. The left panel of Fig.\ \ref{fig-fss} shows the reduced
localization length $\lambda_{\rm 3}/M$ for even $M$ up to $14$ at
$E=0$ and $c=0$.  The transition from extended states for $W \lesssim
4.05$ to localized states for $W \gtrsim 4.05$ is clearly visible as
$\lambda_3/M$ decreases (increases) with increasing $M$ for the
extended (localized) case. In the right graph the resulting scaling
function $f_3(\xi/M)$ is plotted.  It was obtained by a non-linear fit
taking non-linear and non-universal corrections to FSS \cite{SleO99a}
into account.  Comparing the exponents $\nu_E = 1.61\pm 0.07$ and
$\nu_W=1.54\pm 0.03$ of the transitions obtained from the fits at
$W=0$, $c=0$ and $E=0$, $c=0$, respectively, with recent results for
the Anderson model with pure diagonal disorder we find good agreement.

\noindent{\bf Conclusion.}
In this work we have studied the localization properties of the
Anderson model of localization with off-diagonal disorder. We find
non-localized states in 1D and 2D only at the band center.  In quasi
1D we examined odd/even system size effects for the localization
length. We showed that the wave functions at $E=0$ in 2D remain
critical up to the system sizes considered regardless of the odd/even
effects. Our numerical results match reasonably well with the
predictions \cite{InuTA94,BroMF99}.  In the 3D case, we
obtained the critical exponents of the MIT using TMM and FSS.  Their
values agree with recent results for the model with only diagonal
disorder \cite{SleO99a,Mac94}. Thus although the off-diagonal case
exhibits some unusual features, its physics is nevertheless accurately
described by the orthogonal universality class within the scaling
theory of localization \cite{KraM93,AbrALR79}.

\noindent{\bf Acknowledgements.}
We thank C.\ M.\ Soukoulis and J.\ Stolze for pointing out Refs.\
\cite{SouWGE82,InuTA94} to us.

\clearpage
%

\clearpage

\pagestyle{empty}

\begin{figure}
\epsfxsize=\textwidth
\epsfbox{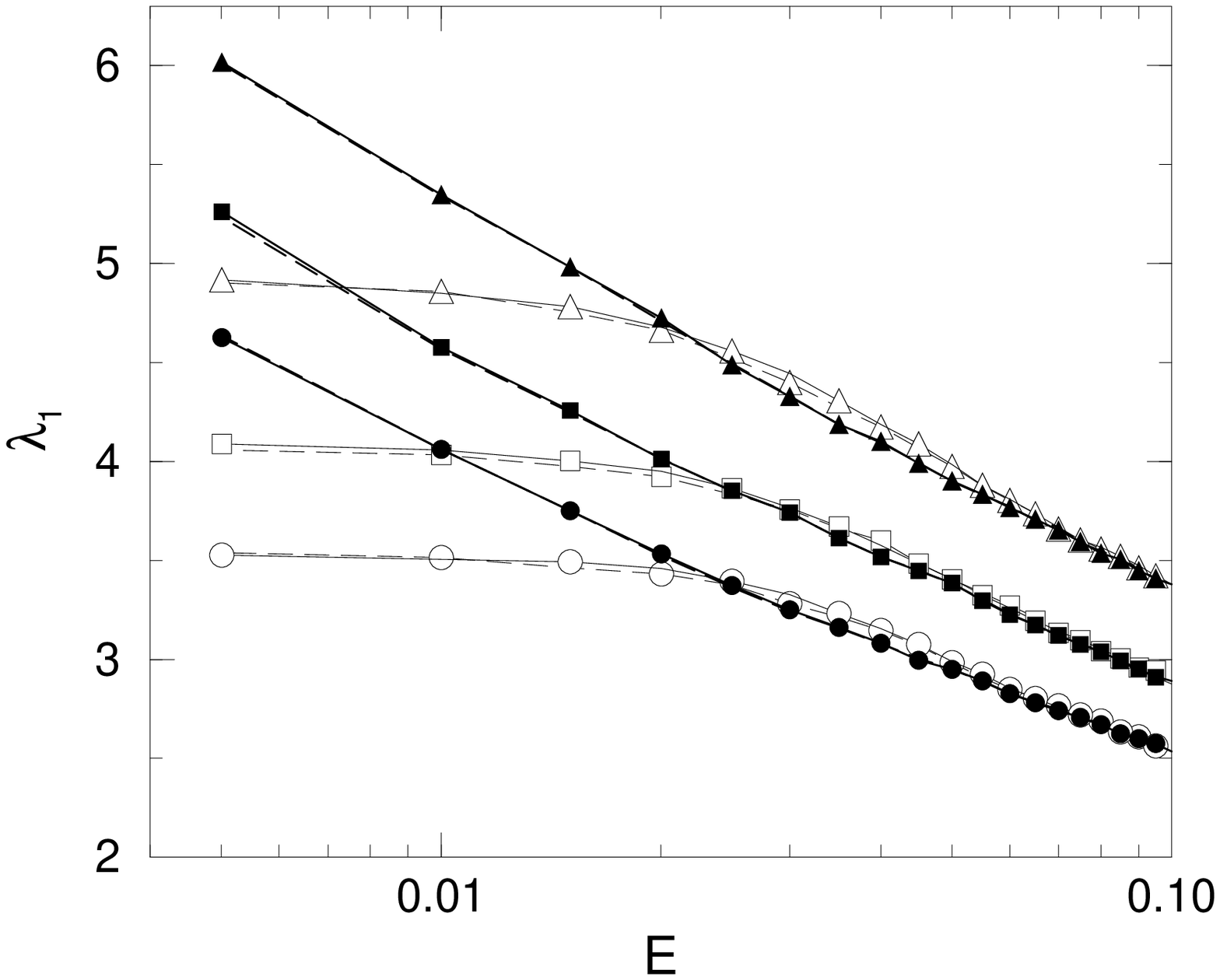}
\end{figure}

\clearpage
\begin{figure}
\epsfxsize=\textwidth
\epsfbox{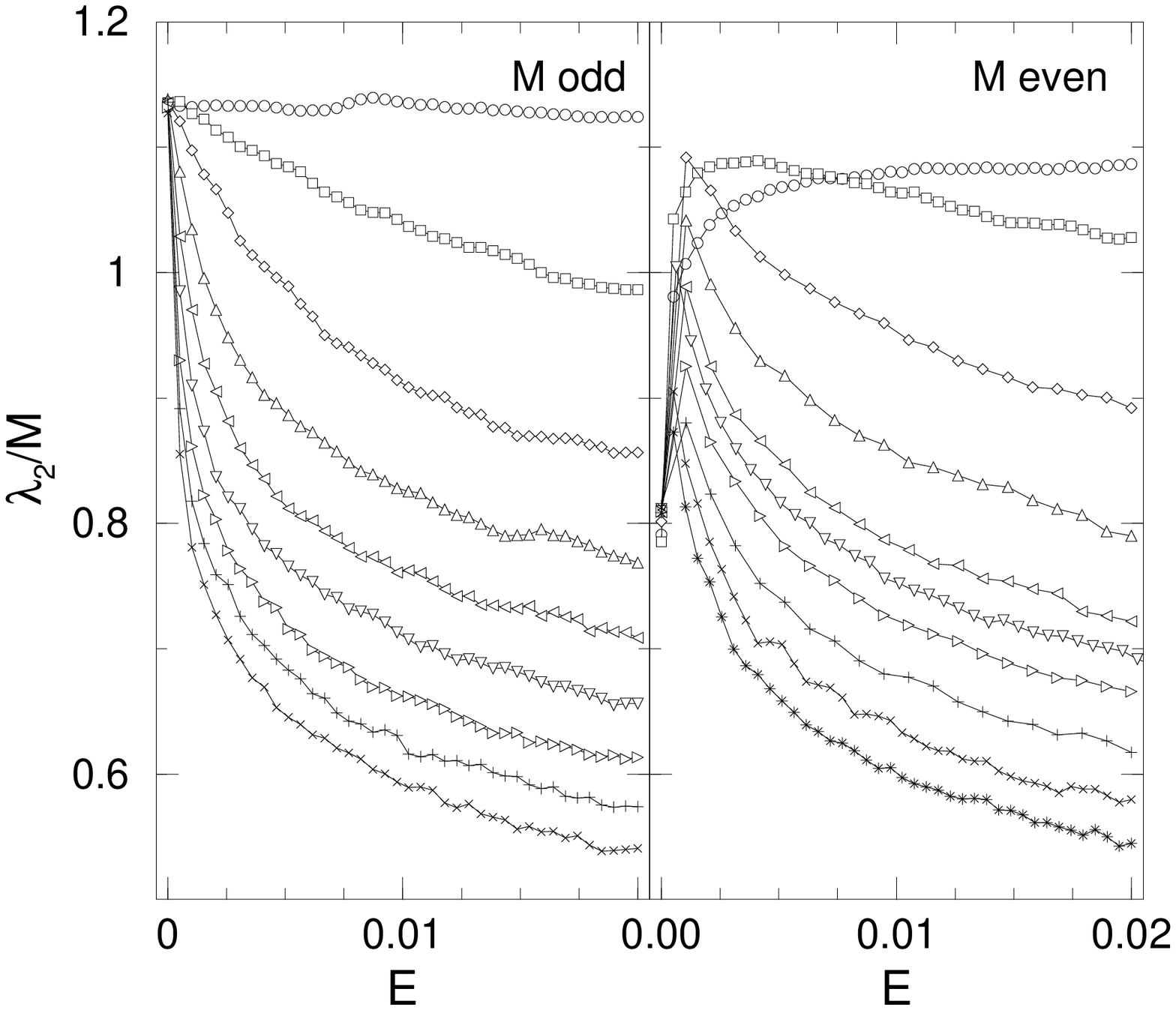}
\end{figure}

\clearpage
\begin{figure}
\epsfxsize=\textwidth
{\epsfbox{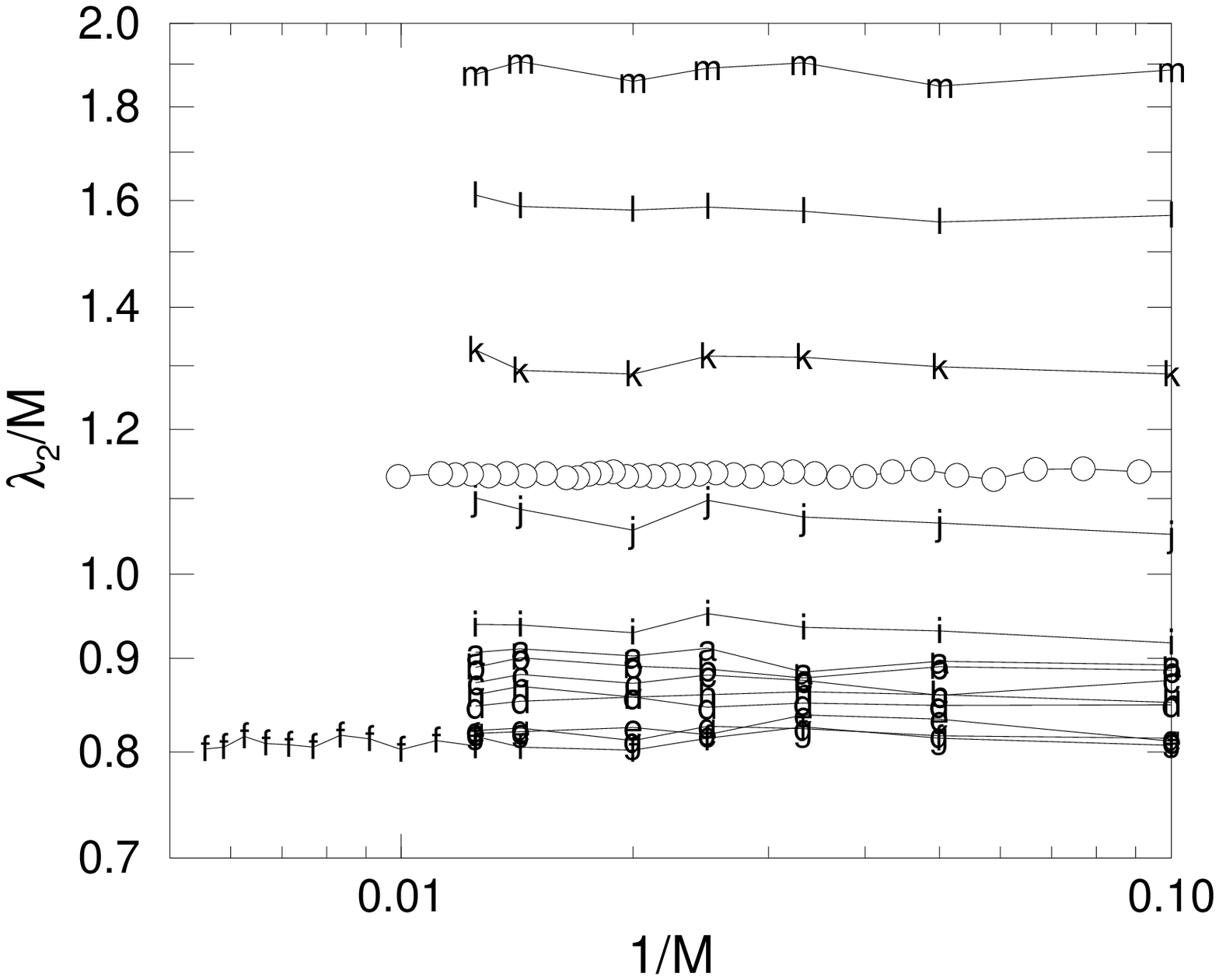}}
\end{figure}

\clearpage
\begin{figure}
\vspace*{8.5cm}
\epsfysize=0.6\textheight
\epsfbox{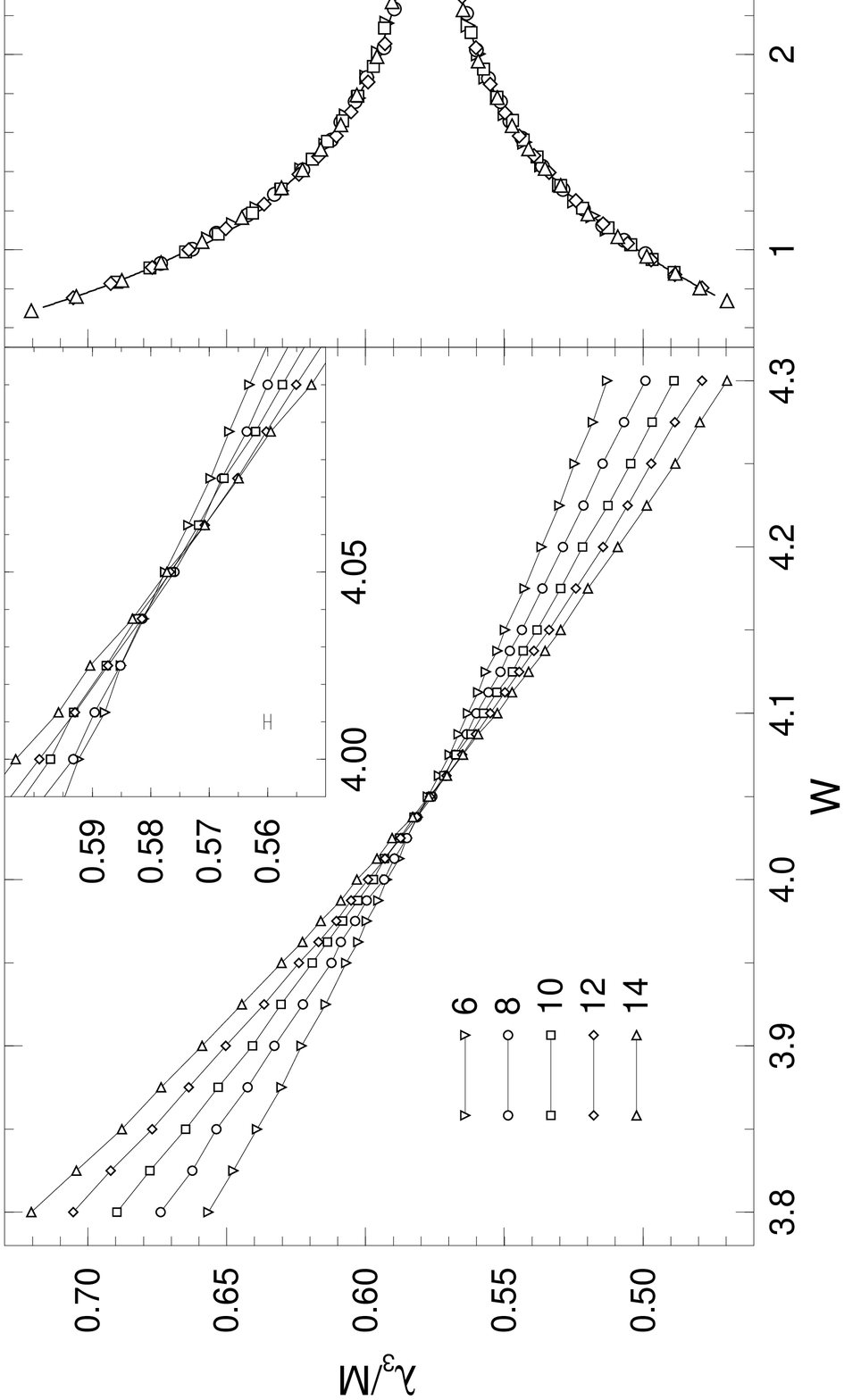}
\end{figure}
\end{document}